\documentclass[useAMS,usenatbib]{mn2e}
\usepackage{graphicx}
\usepackage{color}
\usepackage{ulem}
\def\apj{ApJ}
\def\apjs{ApJS}

\def\aap{A\&A}
\def\aaps{A\&AS}

\def\mnras{MNRAS}

\def\pasp{PASP}

\def\araa{Ann.Rev.Astron.Astrophys.}

\title[HSC J1631$+$4426]
{Optical spectroscopy of the extremely metal-deficient star-forming galaxy
HSC J1631+4426: a test of the strong-line method}
\author[T. X. Thuan et al.]{T. X.\ Thuan$^{1}$, N. G.\ Guseva$^{2}$ and
  Y. I.\ Izotov$^{2}$\thanks{Corresponding author: yizotov@bitp.kiev.ua}\\
                $^{1}$Astronomy Department, University of Virginia, 
                     P.O. Box 400325, Charlottesville, VA 22904-4325,\\
                $^{2}$Bogolyubov Institute for Theoretical Physics,
                     National Academy of Sciences of Ukraine,
                     14-b Metrolohichna str., Kyiv, 03143, Ukraine,\\
}
\begin{document}


\pagerange{\pageref{firstpage}--\pageref{lastpage}} \pubyear{2022}

\maketitle

\label{firstpage}

\begin{abstract}
Recently, Kojima and co-authors have reported a record low oxygen abundance,
12~+~logO/H = 6.90\,$\pm$\,0.03, or 1.6\% of solar metallicity, in the low-mass star-forming galaxy
HSC J1631$+$4426.  This exceptionally low oxygen abundance was obtained by the
direct method, using the [O~{\sc iii}]$\lambda$4363\AA\ emission line. 
However, using the strong-line method by Izotov et al. (2019b), these authors
have derived a significantly higher metallicity 12~+~logO/H = 7.175\,$\pm$\,0.005. 
To clarify the situation, we have obtained new observations of HSC J1631$+$4426
with the Large Binocular Telescope (LBT)/Multi-Object Dual Spectrograph (MODS).
We have derived a higher oxygen abundance, 12~+~logO/H = 7.14\,$\pm$\,0.03, using
the direct method, a value similar to the oxygen abundance obtained by the 
strong-line method. Thus, HSC J1631$+$4426 has a metallicity close to that of
the well known blue compact dwarf galaxy I Zw 18.
\end{abstract}


\section{Introduction}\label{sec:INT}

Extremely metal-deficient (XMD) galaxies with active star formation
constitute a rare and intriguing class of objects in the local Universe. We
shall define here XMD galaxies as those having oxygen abundances
12~+~logO/H~$\leq$~7.3, or 4\% solar, taking the solar value to be 8.7 \citep{asp09}.
Despite the fact that these low-$z$ XMD galaxies have
a gaseous content that is not quite as pristine as that of primordial galaxies,
they do represent their best local counterparts and thus, can be used to
compare with the high-$z$ primeval objects to be observed in the near-future
by the {\sl James Webb Space Telescope} ({\sl JWST}) and the 30 m-class
ground-based telescopes.

In the last few years, a number of nearby galaxies have been discovered with
extremely low oxygen abundances 12~+~logO/H $\sim$ 7.0 (or 2\% solar). Thus, \citet{H16} have
found 12~+~logO/H = 7.02\,$\pm$\,0.03 for the galaxy AGC~198691, and the value of 
7.13\,$\pm$\,0.08 has been derived by \citet{H17} for the dwarf galaxy Little Cub.
\citet{I18a}, \citet{Iz19a} and \citet{IzTGus21} have reported
12~+~logO/H = 6.98\,$\pm$\,0.02,  7.035\,$\pm$\,0.026 and 7.085\,$\pm$\,0.031 for the
galaxies J0811$+$4730,  J1234$+$3901 and J2229$+$2725, respectively.
Recently,  \citet{Kojima2020} have derived a record low oxygen abundance of
6.90\,$\pm$\,0.03 in the XMD galaxy HSC J1631+4426.

\begin{figure*}[t!]
\hbox{
\includegraphics[angle=-90,width=0.42\linewidth]{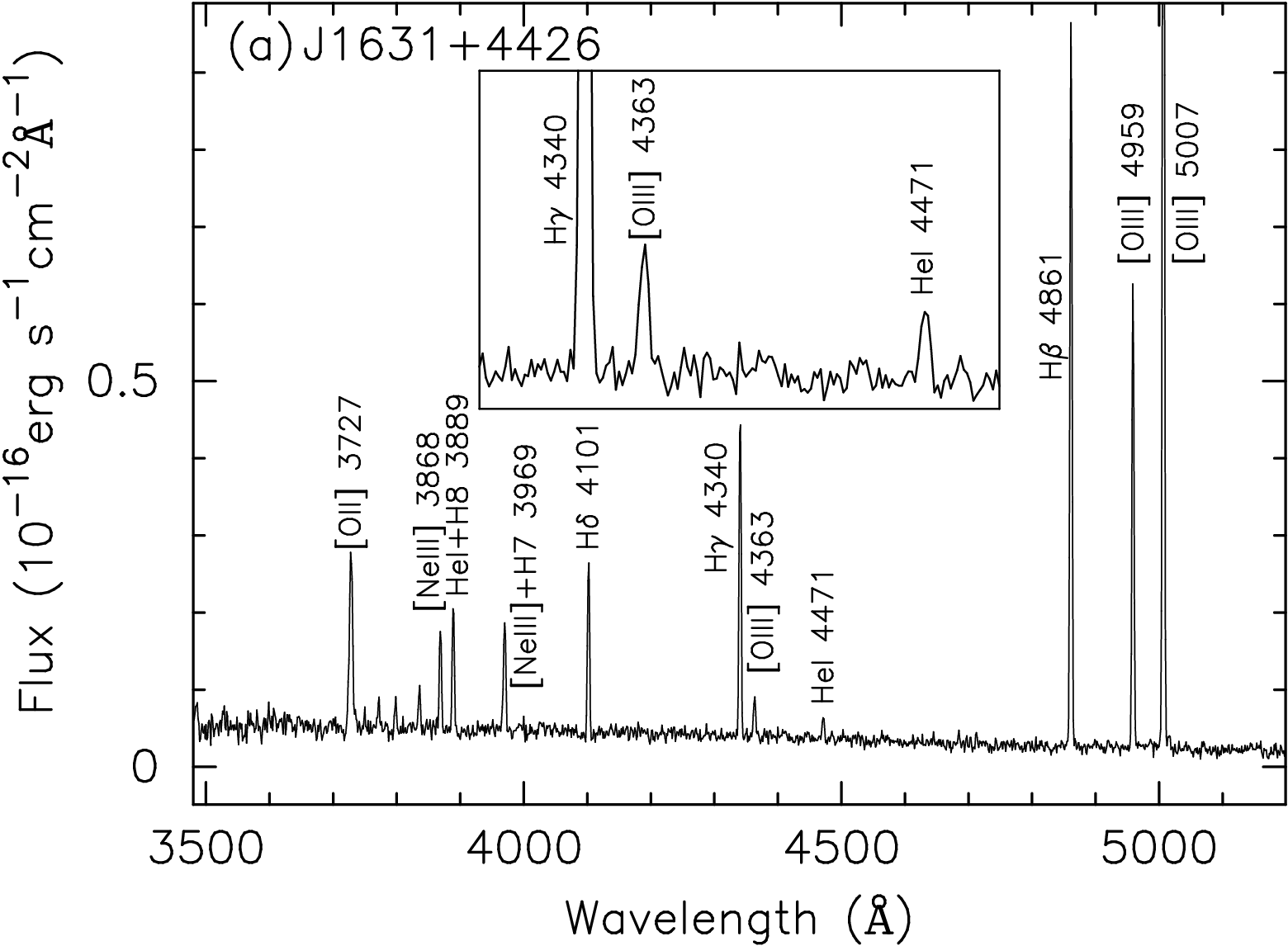}
\hspace{1.0cm}\includegraphics[angle=-90,width=0.42\linewidth]{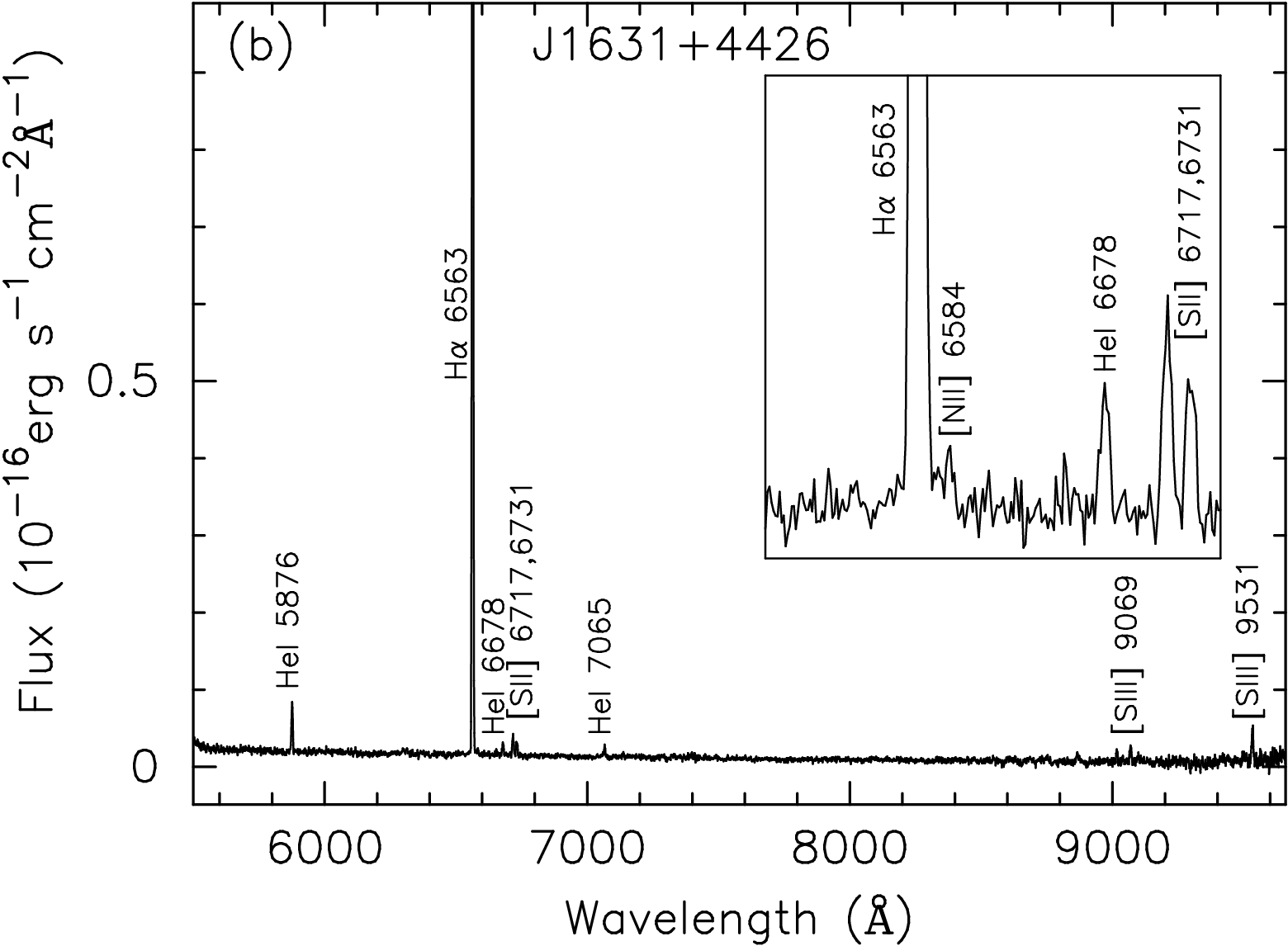}
}
\caption{The rest-frame LBT spectrum of HSC J1631$+$4426. The galaxy's redshift
is 0.03125. Insets in {\bf (a)} and {\bf (b)} show expanded parts of the
spectral regions around the H$\gamma$ and H$\alpha$ emission lines,
respectively, for a better view of weak features. Some emission lines are
labelled. 
}
\label{fig1}
\end{figure*}

  Oxygen abundances in all these compact galaxies have been derived for the
entire galaxy. On the other hand, there have also been cases where extremely low
oxygen abundances have been derived for individual H~{\sc ii} regions in the
same galaxy. Thus, oxygen abundances of 7.01\,$\pm$\,0.07, 6.98\,$\pm$\,0.06,
6.86\,$\pm$\,0.14 have been determined in three H~{\sc ii} regions of the XMD blue
compact dwarf SBS 0335--052W \citep{I09}, while 6.96\,$\pm$\,0.09 has been found
for one of the H~{\sc ii} regions in the dwarf irregular galaxy DDO~68
\citep{An19}.

The most reliable method for abundance determination is the so-called
``direct method''. It requires the detection, with good accuracy, of the
[O~{\sc iii}]$\lambda$4363 emission line which plays the role of an electron
temperature indicator. For star-forming galaxies with a weak or undetected
[O~{\sc iii}]$\lambda$4363 emission line, ``strong-line methods'' are used. As
proposed originally by \citet{Pagel1979}, these methods are based on
combinations of strong emission-line intensities of various elements. The oxygen
abundance indicator R$_{23}$ = ([O~{\sc iii}]$\lambda$4959+$\lambda$5007 +
[O~{\sc ii}]$\lambda$3727)/H$\beta$ suggested by \citet{Pagel1979} has met with
widespread acceptance and use. 

For the strong-line method, grids of photoionization models have been used to
calibrate the relation between the line intensities of strong oxygen lines and
the oxygen abundance. However, different models by different authors give
divergent calibrations, so it is best to base the calibration on oxygen
abundances derived from observations through the direct method. The earliest
calibrations were one-dimensional \citep[e.g., ][]{EP84,mccall85,D86}, i.e.,
the oxygen abundance depends on an unique parameter, R$_{23}$. 

Further development in reducing the scatter and improving the accuracy of the
strong-line method is to introduce a correction for the ionization state of
H~{\sc ii} regions. Indeed, as pointed out by \citet{M91}, the intensities of
the oxygen emission lines depend not only on the metallicity of the H~{\sc ii}
region, but also on its ionization state, which can be parameterized by the
ratio O$_{32}$\,=\,[O~{\sc iii}]$\lambda$5007/[O~{\sc ii}]$\lambda$3727. This
ratio is the observational proxy of the ionization parameter $U$.

Many two-dimensional calibrations (the oxygen abundance depends on R$_{23}$ and
O$_{32}$ or a parameter similar to O$_{32}$) were derived for application to
galaxies in a wide range of metallicities, typically at oxygen abundances
12~+~log\,O/H~$\geq$\,7.4 and O$_{32}$~$\la$~5, and often at the expense
of the accuracy in the abundance determination at the lowest metallicities and
high O$_{32}$ \citep[e.g., ][]{PTh05,Nagao2006}.
Recently, \citet{IzTGus21} have
proposed an improved modification of the empirical strong-line calibration by
\citet{Iz19b}, focussed on the oxygen abundance determination in very metal-poor
galaxies, those in the XMD class with 12 + log O/H $\leq$ 7.3.
It is found that all the XMD galaxies discussed above follow closely the
statistical relation between oxygen abundance and strong line ratios, as derived
by \citet{IzTGus21}. There is nevertheless a striking exception, concerning the
star-forming galaxy HSC J1631$+$4426 \citep{Kojima2020}. With its oxygen
abundance of 6.90\,$\pm$\,0.03 derived by the direct method,
it strongly deviates from the strong-line relation. On the other hand,
\citet{Kojima2020,Kojima2021} using the strong-line method by \citet{Iz19b},
derived a much higher oxygen abundance of 7.175\,$\pm$\,0.005.  
   
As the XMD galaxy HSC J1631$+$4426 has the lowest oxygen abundance ever reported
for a star-forming object, it is crucial to resolve any  possible inconsistency
between the oxygen abundances derived by the direct and strong-line methods,
especially when these are applied in the most extreme metallicity regime. To
this end, we have obtained new spectroscopic observations of that galaxy with
the LBT to detect the [O~{\sc iii}]$\lambda$4363 emission line with a high
signal-to-noise ratio, use the direct method to derive an accurate oxygen
abundance, and compare its value with the one derived by the strong-line method.

\section{Observations and data reduction}\label{sec:observations}

We have obtained LBT long-slit spectrophotometric observations of
HSC J1631$+$4426 on 1 May, 2022 in the twin binocular mode, using the MODS1 and
MODS2 spectrographs. Spectra were obtained in the wavelength range
3200~--~10000\AA\ with a 1.2 arcsec wide slit, resulting in a resolving power
$R$ $\sim$ 2000. The seeing during the observations was 0.6  arcsec.

Eight subexposures of approximately 900 s were obtained in both the blue and
red ranges separately with MODS1 and MODS2, resulting in a total exposure time
of 14048~s in the blue range and of 12252~s in the red range, counting separate
exposures with both spectrographs. The airmass during observations was small,
equal to 1.05. Thus, the effect of atmospheric refraction is small for all
subexposures \citep[see ][]{Filippenko1982}. 

The spectrum of the spectrophotometric standard star BD+33~2642 was obtained
with a 5 arcsec wide slit during the same night, for flux calibration and
correction for telluric absorption in the red part.

Bias subtraction, flat field correction, wavelength and flux calibration were
done with the MODS Basic CCD Reduction package {\sc modsccdred}
\citep{Pogge2019} and {\sc iraf}. After these reduction steps, MODS1 and MODS2
subexposures were co-added and one-dimensional spectra of HSC J1631$+$4426 in
the blue and red ranges were extracted in a 1.2 arcsec aperture along the
spatial axis. These spectra exhibit intense emission lines, including a strong
[O~{\sc iii}]$\lambda$4363 emission line (see Fig. \ref{fig1} and insets
therein).

\section{Heavy element abundances}\label{sec:abundances}

  The observed emission-line fluxes and their errors were measured using the 
{\sc iraf} {\it splot} routine. They were corrected for extinction and
underlying hydrogen stellar absorption, derived iteratively from the observed
decrement of the hydrogen Balmer emission lines, following  \citet*{ITL94}.
In our iterative procedure, the equivalent widths of the underlying stellar
Balmer absorption lines are assumed to be the same for all transitions. The
extinction-corrected fluxes together with the extinction coefficient
$C$(H$\beta$), the observed H$\beta$ emission-line flux $F$(H$\beta$), its
rest-frame equivalent width EW(H$\beta$), and the equivalent width of the Balmer
absorption lines are shown in Table~\ref{tab1}.

In the Table, we have also given similar data obtained by \citet{Kojima2021} for
their spectrum of HSC J1631$+$4426. Comparison between the two sets of fluxes
shows that there is general good agreement for the strong lines. For example,
the flux  differences in the [O~{\sc iii}]$\lambda$4959, 5007 lines are
$\sim$5--8 per cent. The fluxes of all other lines are in agreement within the
1$\sigma$ errors. Exceptions are the H$\alpha$/H$\beta$ flux ratio of
\citet{Kojima2021} which is lower than the recombination value, and their higher
[O~{\sc ii}]$\lambda$3727 (by about 20 per cent), probably due to their adopted
higher extinction. Most relevant to this work, our flux of the
[O~{\sc iii}]$\lambda$4363 emission line is $\sim$30 per cent smaller than the
one obtained by \citet{Kojima2021}. This smaller flux  value will have important
consequences on the derived direct-method oxygen abundance. 

\begin{table}
\caption{Extinction-corrected emission-line flux ratios$^{*}$ \label{tab1}}
\begin{tabular}{lrr} \hline
Line& \multicolumn{2}{c}{HSC J1631$+$4426}   \\ 
 & this paper  &  \citet{Kojima2021}   \\ \hline  
3727.00 [O {\sc ii}]            & 46.34$\pm$1.69 & 50.12$\pm$2.66 \\
3750.15 H12                     &  4.19$\pm$1.28 & ...~~~~~~ \\
3770.63 H11                     &  6.22$\pm$1.08 & ...~~~~~~ \\
3797.90 H10                     &  6.34$\pm$1.00 & ...~~~~~~ \\
3835.39 H9                      &  8.65$\pm$0.84 & 4.68$\pm$1.17\\
3868.76 [Ne {\sc iii}]          & 15.89$\pm$0.75 & 21.73$\pm$1.11 \\
3889.00 He {\sc i}+H8           & 21.84$\pm$0.95 & ...~~~~~~ \\
3968.00 [Ne {\sc iii}]+H7       & 21.32$\pm$0.91 & 20.03$\pm$0.87\\
4101.74 H$\delta$               & 27.37$\pm$1.01 & 27.53$\pm$0.65\\
4340.47 H$\gamma$               & 48.34$\pm$1.54 & 46.88$\pm$0.50\\
4363.21 [O {\sc iii}]           &  6.10$\pm$0.38 & 8.18$\pm$0.48\\
4471.48 He {\sc i}              &  3.25$\pm$0.37 & ...~~~~~~ \\
4686.00 He {\sc ii}             &  1.49$\pm$0.29 & 2.32$\pm$0.38 \\
4711.00 [Ar {\sc iv}]+He {\sc i}&  1.33$\pm$0.29 & $<$0.35~~~ \\
4740.00 [Ar {\sc iv}]           &  0.60$\pm$0.26 & $<$0.36~~~ \\
4861.33 H$\beta$                &100.00$\pm$3.01 & 100.00$\pm$0.37\\
4921.93 He {\sc i}              &  1.23$\pm$0.29 & ...~~~~~~ \\
4958.92 [O {\sc iii}]           & 59.52$\pm$1.81 & 55.76$\pm$0.34\\
4986.00 [Fe {\sc iii}]          &  1.30$\pm$0.36 & ...~~~~~~ \\
5006.80 [O {\sc iii}]           &176.39$\pm$5.24 & 170.92$\pm$0.38\\
5015.68 He {\sc i}              &  1.55$\pm$0.31 & ...~~~~~~ \\
5875.60 He {\sc i}              &  9.24$\pm$0.60 & 9.12$\pm$0.60 \\
6300.00 [O {\sc i}]             &  1.36$\pm$0.40 &  ...~~~~~~ \\
6312.00 [S {\sc iii}]           &  0.96$\pm$0.36 & $<$0.55~~~\\
6562.80 H$\alpha$               &274.98$\pm$8.75 & 229.46$\pm$1.00\\
6583.40 [N~{\sc ii}]            &  1.04$\pm$0.47 & $<$0.48~~~~~~ \\
6678.10 He {\sc i}              &  2.62$\pm$0.52 & 2.07$\pm$0.59 \\
6716.40 [S~{\sc ii}]            &  4.17$\pm$0.55 & ...~~~~~~ \\
6730.80 [S~{\sc ii}]            &  3.00$\pm$0.52 & ...~~~~~~ \\
7065.30 He {\sc i}              &  1.65$\pm$0.57 & ...~~~~~~ \\
7135.80 [Ar~{\sc iii}]          &  0.83$\pm$0.46 & ...~~~~~~ \\
7320.00 [O {\sc ii}]            &  0.58$\pm$0.43 & $<$0.48~~~ \\
7330.00 [O {\sc ii}]            &  0.87$\pm$0.43 & $<$0.54~~~ \\
8750.47 P12                     &  0.76$\pm$0.19 & ...~~~~~~ \\
8862.79 P11                     &  1.72$\pm$0.37 & ...~~~~~~ \\
9014.91 P11                     &  2.01$\pm$0.43 & ...~~~~~~ \\
9069.00 [S {\sc iii}]           &  3.53$\pm$0.68 & ...~~~~~~ \\
9530.60 [S {\sc iii}]           &  6.86$\pm$1.18 & ...~~~~~~ \\
$C$(H$\beta$)$^{\dag}$        &0.145$\pm$0.038 & 0.28$\pm$0.04\\
$F$(H$\beta$)$^{\ddag}$       &3.85$\pm$0.03   & 1.31$\pm$0.04\\
EW(H$\beta$)$^{**}$           &155.3$\pm$1.0  & 123.5$\pm$3.2\\
EW(abs)$^{**}$                &1.8$\pm$0.2    & ...~~~~~~ \\
\hline
  \end{tabular}

Notes: $^{*}$in units 100$\times$$I(\lambda)$/$I$(H$\beta$). $^{\dag}$Extinction coefficient, derived from the observed hydrogen Balmer decrement. $^{\ddag}$Observed flux in units of 10$^{-16}$ erg s$^{-1}$ cm$^{-2}$. $^{**}$Equivalent width in \AA.

  \end{table}

\begin{table}
\caption{Electron temperatures, electron number density 
and heavy element abundances \label{tab2}}
\begin{tabular}{lcc} \hline
  Property& \multicolumn{2}{c}{HSC J1631$+$4426}  \\
 & this paper  &  \citet{Kojima2021}   \\ \hline    
$T_{\rm e}$(O {\sc iii}), K          &   20300$\pm$800  & 25570$\pm$1100 \\
$T_{\rm e}$(O {\sc ii}), K           &   15600$\pm$600  & ...     \\
$T_{\rm e}$(S {\sc iii}), K          &   18900$\pm$700  & ...     \\
$N_{\rm e}$(S {\sc ii}), cm$^{-3}$    &30$\pm$220       & ...     \\ \\
O$^+$/H$^+$$\times$10$^5$            &0.372$\pm$0.038  & ... \\
O$^{2+}$/H$^+$$\times$10$^5$          &0.981$\pm$0.093  & ... \\
O$^{3+}$/H$^+$$\times$10$^6$          &0.232$\pm$0.067 \\
O/H$\times$10$^5$                   &1.376$\pm$0.100   & ... \\
12+log(O/H)                         &7.139$\pm$0.032 & 6.90$\pm$0.03 \\ \\
N$^{+}$/H$^+$$\times$10$^7$          &0.706$\pm$0.025 & ... \\
ICF(N)                              &3.654           & ... \\
N/H$\times$10$^7$                   &2.579$\pm$0.921 & ... \\
log(N/O)                            &$-$1.727$\pm$0.158~ & $<$ $-$1.710\\ \\
Ne$^{2+}$/H$^+$$\times$10$^6$        &1.947$\pm$0.188 & ... \\
ICF(Ne)                             &1.121           & ... \\
Ne/H$\times$10$^6$                  &2.183$\pm$0.264 & ... \\
log(Ne/O)                           &$-$0.803$\pm$0.060~& $-$0.641$\pm$0.020 \\ \\
S$^{+}$/H$^+$$\times$10$^7$          &0.643$\pm$0.077 & ... \\
S$^{2+}$/H$^+$$\times$10$^7$         &2.561$\pm$0.967 & ... \\
ICF(S)                             &1.140           & ... \\
S/H$\times$10$^7$                   &3.653$\pm$1.109 & ... \\
log(S/O)                           &$-$1.576$\pm$0.136~&  ... \\ \\
Ar$^{2+}$/H$^+$$\times$10$^7$          &0.229$\pm$0.125 & ... \\
ICF(Ar)                             &1.0999          & ... \\
Ar/H$\times$10$^7$                   &0.251$\pm$0.200 & ... \\
log(Ar/O)                           &$-$2.739$\pm$0.347~&  ... \\ \\
Fe$^{2+}$/H$^+$$\times$10$^6$          &0.223$\pm$0.064 & ... \\
ICF(Fe)                             &4.9962          & ... \\
Fe/H$\times$10$^6$                   &1.113$\pm$0.321 & ... \\
log(Fe/O)                           &$-$1.092$\pm$0.125~&$-$1.246$\pm$0.220 \\
\hline
  \end{tabular}
  \end{table}

We follow the prescriptions of \citet{I06} to derive the electron
temperature and density and heavy element abundances from extinction-corrected
fluxes of emission lines in HSC J1631$+$4426. The electron temperature
$T_{\rm e}$(O~{\sc iii}) is calculated from the
[O~{\sc iii}]$\lambda$4363/($\lambda$4959 + $\lambda$5007) emission-line flux
ratio. It is used to derive the abundances of O$^{2+}$, O$^{3+}$ and Ne$^{2+}$.
The abundances of O$^{+}$, N$^{+}$, S$^{+}$ and Fe$^{2+}$ are derived with the
electron temperature $T_{\rm e}$(O~{\sc ii}), using the relations of \citet{I06}
between $T_{\rm e}$(O~{\sc ii}) and $T_{\rm e}$(O~{\sc iii}). To derive abundances
of S$^{2+}$ and Ar$^{2+}$ we adopt the relation between $T_{\rm e}$(S~{\sc iii})
and $T_{\rm e}$(O~{\sc iii}) by \citet{I06}. The electron number density was
derived from the [S~{\sc ii}]$\lambda$6717/$\lambda$6731 flux ratio. The
electron temperatures $T_{\rm e}$(O~{\sc iii}), $T_{\rm e}$(O~{\sc ii}) and
$T_{\rm e}$(S~{\sc iii}), and the electron number density $N_{\rm e}$(S~{\sc ii})
are shown in Table~\ref{tab2}. We obtain an electron temperature
$T_{\rm e}$(O~{\sc iii}) = 20300\,$\pm$\,800K, compared to the exceptionally high
electron temperature $T_{\rm e}$(O~{\sc iii}) = 25570\,$\pm$\,1100K obtained by
\citet{Kojima2020}. 

The ionic abundances, ionisation correction factors ($ICF$s) and total O, N,
Ne, S, Ar and Fe abundances are obtained using relations by \citet{I06}. Using
the direct method, we derive 12\,+\,logO/H\,=\,7.139\,$\pm$\,0.032 for HSC J1631$+$4426
(Table~\ref{tab2}), significantly higher than the oxygen abundance
12\,+\,logO/H\,=\,6.90\,$\pm$\,0.03 obtained by \citet{Kojima2020,Kojima2021}. The N/O,
Ne/O, S/O, Ar/O and Fe/O abundance ratios are similar to those in other
star-forming dwarf galaxies. 

To derive element abundances, we have used the prescriptions of \citet{I06} as
they are based on fairly recent atomic data and photoionization models. We have
compared our results with those obtained by using relations based on other
models. There is nearly complete agreement between our results and those
obtained with the \citet{S90} relations. For HSC J1631$+$4426, the difference
in the electron temperature $T_{\rm e}$(O~{\sc ii}) would be only $\sim$25K, thus
giving the same oxygen abundance. \citet{Kojima2020} used the empirical relation
$T_{\rm e}$(O~{\sc ii})\,=\,0.7\,$\times$\,$T_{\rm e}$(O~{\sc iii})\,+\,3000
of \citet{CTM86}. With our derived  $T_{\rm e}$(O~{\sc iii})\,=\,20300K, the
latter relation would give $T_{\rm e}$(O~{\sc ii})\,=\,17210K. Then
12\,+\,logO/H\,=\,7.11, compared to 7.14 derived with the \citet{I06} relations. 

In summary, the difference between the oxygen abundances of HSC J1631$+$4426
derived by  \citet{Kojima2020} and our group comes almost solely from the flux
difference measured for the [O~{\sc iii}]$\lambda$4363 emission line, and not
from the particular relation used to derive $T_{\rm e}$(O~{\sc ii}). Because our
measured [O~{\sc iii}]$\lambda$4363 flux is $\sim$30 per cent smaller than that
of their group,  our derived temperature is lower, and our oxygen abundance is
higher. 

Since there is a discrepancy between the [O~{\sc iii}]$\lambda$4363 fluxes
measured by our two groups, we need to estimate the oxygen abundances in
HSC J1631$+$4426 in another way, distinct  from the direct method. We next
discuss oxygen abundances derived  by the strong-line method in XMD star-forming
galaxies. 

\section{Strong-line method for oxygen abundance determination in XMD galaxies}
\label{SLM}

The direct $T_{\rm e}$ method is by far the most accurate method to determine
oxygen abundances. However, in the case of galaxies with an undetected
[O~{\sc iii}]$\lambda$4363 line,  to determine oxygen abundances, one has to
appeal to strong-line (or indirect) methods based on the fluxes of some of the
brightest lines in the spectra of actively star-forming galaxies. The parameter
R$_{23}$ = ([O~{\sc iii}]$\lambda$4959+$\lambda$5007 +
[O~{\sc ii}]$\lambda$3727)/H$\beta$ is often used. Nonetheless, even using only 
data with [O~{\sc iii}]$\lambda$4363\AA\ measured with an accuracy better than
25\% to calibrate the relation, there remains a rather large scatter in the
relation 12~+~logO/H vs. R$_{23}$.

The next significant step in improving the accuracy of the indirect method for
oxygen abundance determination is the introduction of a correction for the
ionization state of H~{\sc ii} regions, as discussed by \citet{M91}. That
correction can be quantified by the line flux ratio
O$_{32}$\,=\,[O~{\sc iii}]$\lambda$5007/[O~{\sc ii}]$\lambda$3727.
The use of
an ionization correction does indeed reduce the scatter in the relation
by a factor of $\ga$2 \citep{Iz19b}.
Calibrations of the strong-line method have been performed by many authors over
the years \citep[e.g., ][]{Pilyugin2000, PTh05, KD02, Nagao2006,Curti17}.

     \begin{figure}
       \centering
\includegraphics[angle=-90,width=0.75\linewidth]{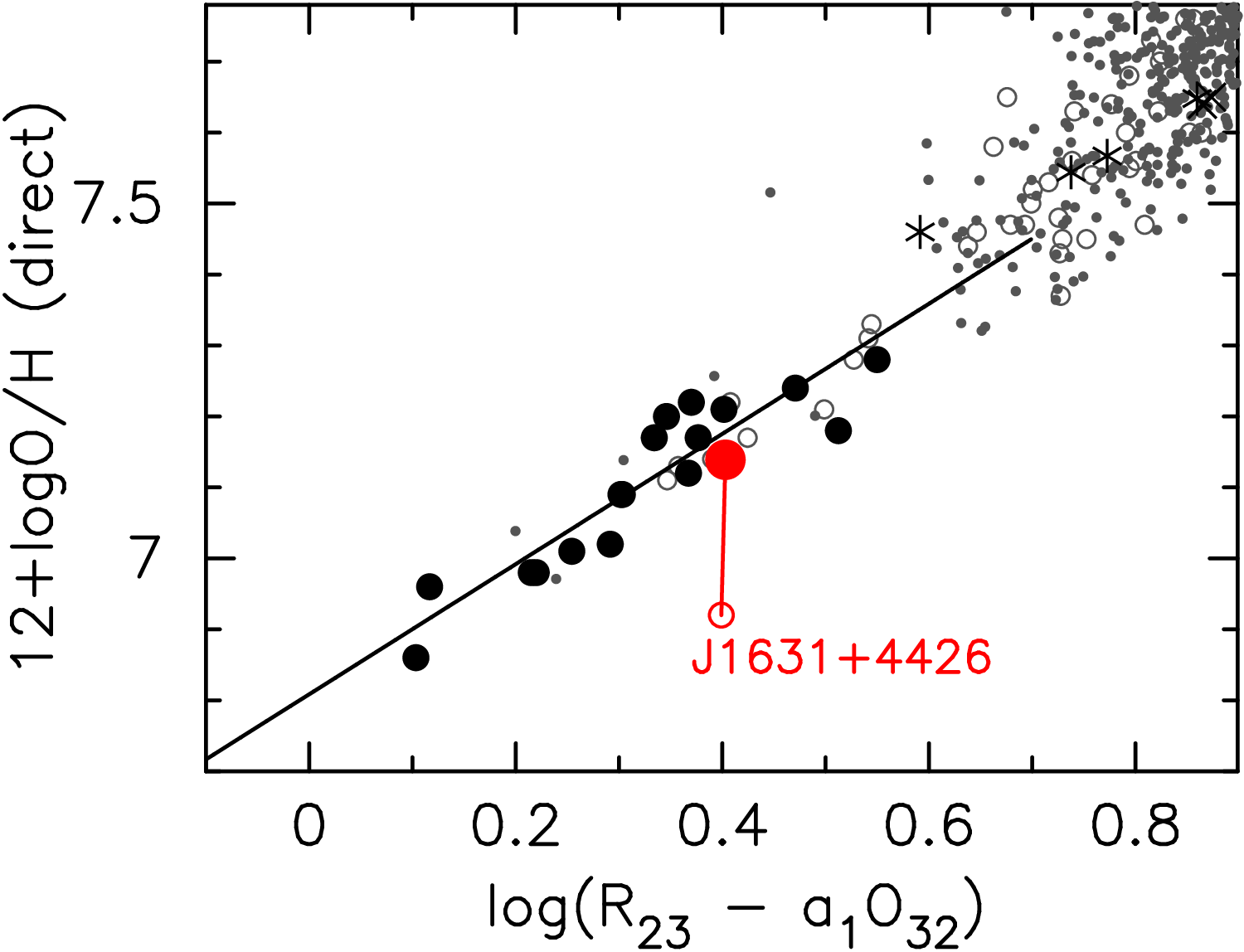}
\caption{
The relation 12\,+\,logO/H\,=\,0.950\,log(R$_{23}$\,--\,$a_1$O$_{32}$)\,+\,6.805
(solid line), where $a_1$\,=\,0.080\,--\,0.00078\,O$_{32}$. The XMDs are
represented by large black filled circles together with the HeBCD sample
(grey open circles), SFGs with O$_{32}$\,$\sim$\,20--40 (grey asterisks) and SFGs
from the SDSS (grey dots) \citep{IzTGus21}.
Locations of HSC J1631$+$4426 using the data from \citet{Kojima2021} and from
this paper are shown by red open and filled circles, respectively.
For all these galaxies 12\,+\,logO/H is derived by the direct method.}
\label{fig2}
\end{figure}

One of the most recent calibrations is that of
\citet{IzTGus21} who derive a new empirical relation for the strong-line
method, specifically tailored for XMD galaxies. It has the form
12\,+\,logO/H\,=\,0.950\,log(R$_{23}$\,--\,$a$$_1$\,O$_{32}$)\,+\,6.805, 
where $a$$_1$\,=\,0.080\,--\,0.00078\,O$_{32}$. 
The scatter of the data points about this relation, which gives an idea of the
abundance uncertainties, is $\sim$ 0.07 dex (Fig.~\ref{fig2}).  

We now discuss the oxygen abundance of HSC J1631$+$4426, for which we present
new observations here. If the oxygen abundance 12~+~logO/H = 6.90\,$\pm$\,0.03 of 
\citet{Kojima2020}, derived by the direct method, is used, then the galaxy is
strongly deviant from the mean 12~+~logO/H vs. log(R$_{23}$ --  $a$$_1$ O$_{32}$)
relation (open red circle in Fig.~\ref{fig2}). However the agreement is
considerably better if the value derived by the direct method from our LBT
observations, 12~+~logO/H = 7.14 \,$\pm$\,0.03, is adopted (filled red circle in
Fig.~\ref{fig2}). We note that the latter oxygen abundance 
is nearly the same as the values of 7.18 and 7.14 obtained respectively by using
the strong-line calibration for XMD galaxies of \citet{IzTGus21}, 
and the strong-line $P$ method of \citet{Pilyugin2000} ($P$ is a quantity
related to O$_{32}$).

The good agreement between the oxygen abundances derived by the direct and
indirect methods gives us confidence both in the accuracy of our direct method
of oxygen abundance determination and in the reliability of the strong-line
method. If our measurement 12~+~logO/H\,=\,7.14\,$\pm$\,0.03 is correct, then the
metallicity of HSC J1631$+$4426 is close to that of the well-known blue compact
dwarf galaxy I~Zw~18. 
The fact that intensive
many-decades long searches for XMD star-forming galaxies have uncovered no
object with a metallicity below 12~+~logO/H $\sim$ 7.0, either in the ionized
gas or neutral gas component \citep*[e.g. ][]{T05} suggests a previous
enrichment of the primordial gas to that oxygen abundance level, perhaps by
Population III stars. This suggestion is supported by observations of
Ly$\alpha$ absorbers which show similar oxygen abundances,
12~+~logO/H $\sim$ 7.0. 

\begin{figure}
\centering{
\includegraphics[angle=-90,width=0.80\linewidth]{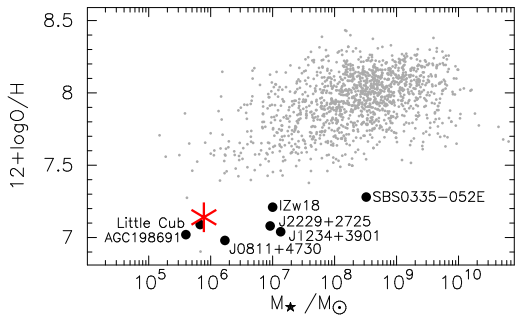}
}
\caption{The stellar mass - metallicity relation. Compact star-forming galaxies
from the SDSS DR14 are represented by grey dots,
XMD galaxies are shown by labelled black filled circles.
HSC J1631$+$4426 is shown by a red asterisk.}
\label{fig3}
\end{figure}

Fig.~\ref{fig3} displays the position of  HSC J1631$+$4426 in the 12~+~logO/H
vs. $M_\star$ plane for XMD galaxIes. The Figure shows that HSC J1631$+$4426 has
properties similar to that of other XMD galaxies.

\section{Summary}\label{sec:conclusions}

We have carried out spectroscopic observations of the low-mass star-forming
galaxy HSC J1631$+$4426 with the Large Binocular Telescope (LBT)/Multi-Object
Dual Spectrograph (MODS). \citet{Kojima2020} have reported a record lowest
oxygen abundance 12~+~logO/H = 6.90\,$\pm$\,0.03 for this galaxy, using the direct
method. However, this value is considerably smaller than the one derived by the
indirect strong-line method. 

From our new observations, we have obtained a higher oxygen abundance,
12~+~logO/H = 7.14\,$\pm$\,0.03 for HSC J1631$+$4426, using the direct method. This
value is also the same as the one obtained by the strong-line method, 
suggesting it is likely to be the correct one.

\section*{Acknowledgements}

N.G.G. and Y.I.I. acknowledge support from the National Academy of Sciences of 
Ukraine by its priority project No. 0122U002259.
``Fundamental properties of the matter and its manifestation
in micro world, astrophysics and cosmology''.
The Large Binocular Telescope (LBT) 
is an international collaboration among institutions in the United States, 
Italy and Germany.
This paper used data obtained with the LBT/MODS 
spectrographs built with
funding from National Science Foundation (NSF) grant AST-9987045 and the NSF Telescope System
Instrumentation Program (TSIP), with additional funds from the Ohio
Board of Regents and the Ohio State University Office of Research.
{\sc iraf} is distributed by the 
National Optical Astronomy Observatories, which are operated by the Association
of Universities for Research in Astronomy, Inc., under cooperative agreement 
with the National Science Foundation.

\section*{Data availability}

The data underlying this article will be shared on reasonable request to the 
corresponding author.

\bsp

\label{lastpage}

\end{document}